\documentclass[aps,prl,superscriptaddress,twocolumn]{revtex4-1}

\usepackage{graphicx,graphics,epsfig}   
\usepackage{dcolumn}    
\usepackage{bm}         
\usepackage{amsmath}    
\usepackage{verbatim}   
\usepackage{color}      
\usepackage{subfigure}  
\usepackage{amsmath,amsfonts,amssymb,graphics,graphics,color,bbm}
\usepackage[rgb]{xcolor}
\usepackage{enumerate}

\newcommand{\tr}{\mathrm{tr}}
\newcommand{\ket}[1]{|#1\rangle}
\newcommand{\bra}[1]{\langle#1|}

\newcommand{\ie}{{\it{i.e.~}}}

\newcommand{\rA}{\mathrm{A}}
\newcommand{\rB}{\mathrm{B}}
\newcommand{\rC}{\mathrm{C}}

\newcommand{\id}{\mathbbm{1}}
\newcommand{\op}{\mathbbm{O}}

\begin{document}

\title{Postquantum steering}

\author{Ana Bel\'en Sainz}\affiliation{H. H. Wills Physics Laboratory, University of Bristol$\text{,}$ Tyndall Avenue, Bristol, BS8 1TL, United Kingdom}
\author{Nicolas Brunner}\affiliation{D\'epartement de Physique Th\'eorique, Universit\'e de Gen\`eve, 1211 Gen\`eve, Switzerland}
\author{Daniel Cavalcanti}\affiliation{ICFO-Institut de Ciencies Fotoniques, Mediterranean Technology Park, 08860 Castelldefels, Barcelona, Spain}
\author{Paul Skrzypczyk}\affiliation{ICFO-Institut de Ciencies Fotoniques, Mediterranean Technology Park, 08860 Castelldefels, Barcelona, Spain}
\author{Tam\'as V\'ertesi}
\affiliation{Institute for Nuclear Research, Hungarian Academy of Sciences,
H-4001 Debrecen, P.O. Box 51, Hungary}
\affiliation{D\'epartement de Physique Th\'eorique, Universit\'e de Gen\`eve, 1211 Gen\`eve, Switzerland}
\begin{abstract}
The discovery of postquantum nonlocality, i.e. the existence of nonlocal correlations stronger than any quantum correlations but nevertheless consistent with the no-signaling principle, has deepened our understanding of the foundations quantum theory. In this work, we investigate whether the phenomenon of Einstein-Podolsky-Rosen steering, a different form of quantum nonlocality, can also be generalized beyond quantum theory. While postquantum steering does not exist in the bipartite case, we prove its existence in the case of three observers. Importantly, we show that postquantum steering is a genuinely new phenomenon, fundamentally different from postquantum nonlocality. Our results provide new insight into the nonlocal correlations of multipartite quantum systems.
\end{abstract}

\maketitle

Quantum mechanics allows for distant systems to be entangled, that is, correlated in a way that admits no equivalent in classical physics. The strongest demonstration of this phenomena is quantum nonlocality \cite{bell64,review}. Performing well-chosen local measurements on separated entangled quantum systems, allows one to observe correlations stronger than in any physical theory satisfying a natural notion of locality, as discovered by Bell. A third form of quantum inseparability is Einstein-Podolsky-Rosen (EPR) steering, which captures the fact that by making a measurement on half of an entangled pair, it is possible to remotely `steer' the state of the other half. First discussed by Schrodinger \cite{schrodinger}, this notion was extensively studied in the context of quantum optics \cite{reid08}. Following a quantum information approach, the concept was put on firm grounds only a few years ago \cite{wiseman07}, and has attracted growing attention since then. The detection \cite{cavalcanti09,saunders10} and quantification \cite{pusey13,skrzypczyk13} of steering have been discussed. The concept was also shown to be relevant in quantum information \cite{cyril,piani}, and related to fundamental aspects of quantum theory such as incompatibility of measurements \cite{uola14,quintino14}.

These phenomena are today viewed as fundamental aspects of quantum theory. Hence a deeper understanding of them provides a fresh perspective on the foundations of quantum theory. In particular, the development of a generalized theory of nonlocality, independent of quantum theory, has brought substantial progress. In a seminal paper, Popescu and Rohrlich discovered the existence of correlations that are stronger than those of quantum theory, but nevertheless satisfying the no-signaling principle, hence avoiding a direct conflict with relativity \cite{PR}. This naturally raised the question of whether there exist physical principles (stronger than no-signaling) from which the limits of quantum nonlocality can be recovered. Significant progress has been reported \cite{popescu14}, notably the discovery of simple information-theoretic and physical principles partly capturing quantum correlations \cite{vanDam,IC,ML,GYNI,LO,ICvsML,almostQ}, and novel derivations of quantum theory based on alternative (arguably more physical) axioms \cite{masanes}. In parallel, this research has led to the device-independent approach, a novel paradigm for ``black-box'' quantum information processing \cite{BHK05,acin07}.

In the present work, motivated by the insight that the study of post-quantum nonlocality has brought, we ask whether the phenomenon of steering can be generalized beyond quantum theory (like nonlocality can), but nevertheless in accordance with the no-signaling principle. We start by discussing the case of two observers (where one party, Bob, steers the other, Alice). Here a celebrated theorem by Gisin \cite{gisin} and Hughston, Josza and Wootters \cite{HJW} implies that postquantum steering does not exist. We then move to the multipartite case, where it is a nontrivial question whether postquantum steering is possible \cite{fritz}. We show explicitly that postquantum nonlocality already implies the existence of postquantum steering when three observers are involved. 
This brings us to the main question and result of the paper, whether postquantum steering that is fundamentally distinct from postquantum nonlocality exists. We discuss what precisely would constitute such a phenomenon, and show that indeed postquantum nonlocality and postquantum steering are genuinely distinct phenomena.

Our results motivate the study of the latter as a new way to study the structure and limitations of quantum correlations. Indeed, the use of the concept of steering allows us to investigate quantum correlations while keeping the local structure of quantum theory. Notably, our results highlight the fact that the structure of the Hilbert space describing tripartite quantum systems is fundamentally different compared to the bipartite case, in accordance with previous work \cite{acinO,toni}.

\emph{Steering in bipartite scenario.--}We start by discussing steering in quantum theory, considering two distant observers, Alice and Bob, sharing a quantum state $ \rho_{\rA\rB}$. Bob wants to convince Alice (who does not trust him) that $ \rho_{\mathrm{AB}}$ is entangled. In order to be convinced, Alice asks Bob to perform various measurements on his system, and to announce the result. Alice can then characterize the state in which her system is steered to, for each measurement of Bob. Although Alice does not know which measurement Bob really performed, she can nevertheless convince herself of the presence of entanglement \cite{wiseman07}.

The conditional (unnormalised) states of Alice's subsystem (prepared by Bob's measurement) are given by
\begin{align}\label{assemblage}
\sigma_{b|y} &= \tr_\rB\left[ \rho_{\rA\rB} \, (\openone_\rA\otimes E_{b|y} ) \right], 
\end{align}
where $E_{b|y}$ denotes the POVM element (effect operator) of Bob corresponding to the outcome $b$ of the measurement setting $y$. Note that $\tr\left[ \sigma_{b|y} \right]$ gives the conditional probability for Bob to obtain outcome $b$ given that he measured $y$, i.e. $p(b|y)$.
The set of unnormalised conditional states $\{\sigma_{b|y}\}_{by}$ is called an \textit{assemblage}. Since any valid POVM satisfies $\sum_b E_{b|y} = \openone$, we have that $\sum_b \sigma_{b|y} = \tr_\rB(\rho_{\rA\rB} )= \rho_\rA$. This can be seen as a statement of the no-signalling principle, since without the knowledge of Bob's outcome $b$, Alice's state is independent of the choice of measurement $y$, being equal simply to the reduced state $\rho_\rA$.

In this work we would like to extend steering beyond quantum theory, and will thus not assume its entire structure. We consider that Alice's system is quantum. Moreover, we assume that the no-signalling principle holds. Thus we are interested in the class of `no-signaling assemblages' which satisfy
\begin{subequations}\label{NSbip}
\begin{align}
\label{PSD_bip} \sigma_{b|y} &\geq 0 &\forall b,y\\
\label{NS-bip} \sum_b \sigma_{b|y} &= \sum_b \sigma_{b|y^\prime} = \rho_\mathrm{A} &\forall y,y' \\
\label{norm_bip} \tr(\rho_\mathrm{A})  &= 1.
\end{align}
\end{subequations}
The first constraint says that Alice's systems is described by (unnormalised) quantum states, i.e. positive semidefinite matrices, the second says that the assemblage should satisfy the no-signalling constraint, and the last that the reduced state of Alice should be normalized.

The question we are interested in is whether every no-signaling assemblage admits a quantum realisation. That is, for any $\{\sigma_{b|y} \}_{by}$ satisfying the constraints \eqref{NSbip}, can we find a set of POVMs $E_{b|y}$ and a quantum state $\rho_{\rA\rB}$ such that $\sigma_{b|y} = \tr_\mathrm{B}\left[ \rho_{\mathrm{AB}} \, \openone_\mathrm{A} \otimes E_{b|y}  \right]$. In other words, can Alice test whether Bob is using postquantum resources to prepare the assemblage.

It turns out that in the bipartite case, every no-signaling assemblage admits a quantum realization. Hence, there is no postquantum steering in this case. This follows from the GHJW theorem \cite{gisin,HJW}, which gives an explicit quantum realization. Given a no-signaling assemblage, condition (\ref{PSD_bip}) implies that $\rho_A$ is positive semidefinite, and hence can be diagonalised: $\rho_\rA = \sum_k \mu_k \ket{k} \bra{k}$. Now define the quantum state $\ket{\Psi}_{\rA\rB} = \sum_k \sqrt{\mu_k} \ket{k}_\rA \ket{k}_\rB$, (in the Schmidt form) and POVM element for Bob $E_{b|y} =  \sqrt{\rho_\rA^{-1}} \, \sigma^{\mathrm{T}}_{b|y} \, \sqrt{\rho_\rA^{-1}}$, where $\sqrt{\rho_\rA^{-1}} = \sum_k 1/\sqrt{\mu_k} \ket{k}_\rA \bra{k}$. It can be checked that $\ket{\Psi}_{\rA\rB}$ is a normalised state, that $\{E_{b|y}\}_b$ is a well defined POVM for each $y$, and that the assemblage is recovered, \ie  $\sigma_{b|y} = \tr_\rB  \left( \ket{\Psi}\bra{\Psi}_{\rA\rB}  \openone_\mathrm{A} \otimes E_{b|y} \right)$.

Below we will show that the situation is completely different in the multipartite case. Specifically, there exist tripartite assemblages which satisfy the no-signaling principle yet admit no quantum realisation.

\emph{Steering in the tripartite scenario.--} Quantum steering has been recently discussed in the multipartite case \cite{he,Cav15}. Following the approach of Ref. \cite{Cav15}, we discuss a tripartite steering scenario where only one observer (Alice) is trusted (characterised). Consider a tripartite quantum state $\rho_{\mathrm{ABC}}$ shared between Alice, Bob and Charlie, and let Bob and Charlie perform (uncharacterized) POVMs $E_{b|y}$ and $E_{c|z}$ on their subsystems. In this case, the assemblage (i.e. the set of unnormalised states for Alice's system) is given by \cite{footnote1}
\begin{align}\label{quant_tri}
\sigma_{bc|yz} &= \tr_\mathrm{BC}\left[ \rho_{\mathrm{ABC}} \, (\openone_\mathrm{A} \otimes E_{b|y} \otimes E_{c|z} ) \right] & \forall b,c,y,z.
\end{align}
Similarly to above, we have that $p(bc|yz) = \tr\left( \sigma_{bc|yz} \right)$. Moreover, no-signalling is ensured, since $\sum_b \sigma_{bc|yz} = \sum_b \sigma_{bc|y^\prime z}$ and $\sum_c \sigma_{bc|yz} = \sum_c \sigma_{bc|yz^\prime}$ $\forall y,y',z,z'$. Finally, Alice's  reduced state is $\sum_{bc}\sigma_{bc|yz} = \rho_\mathrm{A}$.

Again, we would like to extend steering beyond quantum theory, and consider assemblages limited only by the no-signaling principle.
Thus, we are interested in the set of no-signaling assemblages $\sigma_{bc|yz}$ that satisfy
\begin{subequations}\label{NS trip}
\begin{align}
\label{PSD_tri}\sigma_{bc|yz} &\geq 0 &\forall b,c,y,z \\
\label{red_i}\sum_b \sigma_{bc|yz} &= \sum_b \sigma_{bc|y^\prime z} = \sigma_{c|z}^\mathrm{C} &\forall y,y',c,z\\
\label{red_f}\sum_c \sigma_{bc|yz} &= \sum_c \sigma_{bc|yz^\prime} = \sigma_{b|y}^\mathrm{B} &\forall b,y,z,z'\\
\label{norm_tri}\tr\sum_{bc} \sigma_{bc|yz} &= \tr (\rho_\mathrm{A}) = 1
\end{align}
\end{subequations}
where the first constraint imposes positivity, the second no-signalling from Bob to Alice-Charlie, the third no-signalling from Charlie to Alice-Bob, and the fourth normalisation.

We will now show that, contrary to the bipartite case, there exist no-signaling assemblages (i.e. satisfying conditions \eqref{NS trip}) which do not admit a quantum realisation (i.e. cannot be written in the form \eqref{quant_tri}). Hence postquantum steering is possible in the tripartite case. We will first present a simple example which demonstrates that postquantum steering is trivially implied by the existence of postquantum nonlocality. We will then move on to the much more interesting question, namely the existence of postquantum steering that is not implied by postquantum non-locality, i.e. that is fundamentally different from it.

Consider first an assemblage for which the behaviour of Bob and Charlie, $p(bc|yz)$, is not realizable in quantum theory \cite{footnote2}, for instance the PR-box correlations \cite{PR}: $p(bc|yz) = 1/2$ if $b\oplus c = yz$ and $0$ otherwise, with uniform marginals, and where $y,z,b,c=0,1$. Then take any normalised positive semidefinite operator $\rho_\mathrm{A}$ and define $\sigma_{bc|yz} = p(bc|yz) \, \rho_\mathrm{A} $. Clearly, this assemblage is no-signaling, but cannot be realized in quantum theory. This is thus an example of postquantum steering. However, in this (extreme) example Alice is completely factorised from Bob and Charlie, and the postquantumness follows only from the untrusted parties -- i.e. it follows already at the level of nonlocality. Thus examples of this type are not insightful, since they don't rely on the fact that one party is trusted.

\emph{Post-quantum steering without post-quantum nonlocality.}--We are now ready to discuss our main result, namely the existence of post-quantum steering which does not reduce to postquantum nonlocality. At this point it is useful to discuss what exactly would constitute a non-trivial example of postquantum steering. In the previous example we saw that the postquantumness of the assemblage involving a PR-box could be certified directly from the nonlocal behaviour of the untrusted devices, i.e. by tracing out the trusted party. One possibility this suggests is therefore to look for those assemblages $\sigma_{bc|yz}$ such that $p(bc|yz) = \tr[\sigma_{bc|yz}]$ are quantum. This however is not the strongest requirement we could ask for, since it neglects the trusted party altogether. We could still ask the trusted party to measure it's assemblage, using a set of POVMs $E_{a|x}$, to produce the tripartite behaviour $p(abc|xyz) = \tr[E_{a|x}\sigma_{bc|yz}]$. If this behaviour is postquantum for some well chosen set of $E_{a|x}$ then the postquantumness of the assemblage can be witnessed at the level of the nonlocal behaviour it produces. Therefore what we are looking for is an assemblage such that no matter what set of measurements Alice performs she will always produce behaviours explainable within quantum mechanics, yet which is nevertheless post-quantum at the level of the assemblage itself.

In what follows we will outline a method to find an assemblage $\sigma_{bc|yz}$ which: (i) is provably post-quantum (ii) for all POVMs $E_{a|x}$ (with $x$ now a continuous label), the resulting tripartite behaviour $p(abc|xyz) = \tr_\mathrm{A}(E_{a|x}\sigma_{bc|yz})$ admits a quantum realization. The example assemblage with these properties will be a collection of (real) qutrit states. To arrive at our example we will first construct a qubit assemblage which is local for the restricted set of projective measurements. This can then be used to construct a qutrit assemblage which is local for all POVMs.

\textit{Outline of method.---} The first ingredient we need is a test for certifying that a given assemblage is postquantum, i.e. cannot be written in the form \eqref{quant_tri}. To do so we can use so-called \emph{Tsirelson bounds} \cite{Tsirelson} for steering inequalities. Consider a linear steering functional 
\begin{equation}\label{ineq}
\beta = \tr\Big( \sum_{bcyz} F_{bcyz} \, \sigma_{bc|yz} \Big).
\end{equation}
for given operators $\{F_{bcyz}\}_{bcyz}$ \cite{pusey13}. The Tsirelson bound $\beta_\mathcal{Q}$ for this functional is the minimum possible value that can be obtained by assemblages which arise from measurements on quantum states. Hence if a given assemblage $\sigma_{bc|yz}$ is such that $\tr( \sum_{bcyz} F_{bcyz}\sigma_{bc|yz}) = \beta < \beta_\mathcal{Q}$, we can conclude that $\sigma_{bc|yz}$ is post-quantum. The problem with this is that calculating the Tsirelson bound $\beta_Q$ of a steering functional is in general a hard problem, since there is no efficient characterisation of the set of quantum assemblages \cite{TorNavVer14}. However, it is possible to lower bound the Tsirelson bound, $\beta_\mathcal{\widetilde{Q}} \leq \beta_\mathcal{Q}$, in a computationally feasible way, inspired from methods used in the context of quantum nonlocality \cite{NPA}. Full details of how this can be done can be found in the supplementary material A.

The second ingredient needed is a method for constructing assemblages $\sigma_{bc|yz}$ that always produce behaviours which admit quantum realizations for projective measurements, that is such that $p(abc|xyz) = \tr_\mathrm{A}(\Pi_{a|x}\sigma_{bc|yz})$ admits a quantum realization for all possible projective measurements $\Pi_{a|x}$ performed by Alice. Here the challenge arises from the fact that $x$ runs over a continuous set. Nevertheless, inspired by \cite{Bowles2015}, the problem can be reduced to finding a quantum realization for only a finite set of fixed POVMs. First of all, we will make our requirement even more stringent: that the behaviours arising from the assemblage admit a \emph{local} model, and not just a quantum realization, since the set of local behaviours is contained inside the set of quantum behaviours and is easier to characterise \cite{review}. Second, we will use two observations: (i) that noisy POVMs of the form $\Pi_{a|x}(\mu) = \mu \Pi_{a|x} + (1-\mu) \openone/2$ produce the same behaviour on the assemblage $\sigma_{bc|yz}$ as (noiseless) projective measurements do on the noisy assemblage $\sigma_{bc|yz} (\mu)= \mu \sigma_{bc|yz} + (1-\mu) \tr[ \sigma_{bc|yz}] \openone/2$,
\begin{equation}
\tr[ \Pi_{a|x}(\mu) \sigma_{bc|yz} ]= \tr[ \Pi_{a|x} \sigma_{bc|yz}(\mu) ].
\end{equation}
That is, the simulation of noisy measurements on a noiseless assemblage is equivalent to the simulation of noiseless measurements on a noisy assemblage.
(ii) the set of noisy measurements such that $\mu < 1$ can be simulated by a finite set of projective measurements $\mathcal{E}$ \cite{footnote3}. Thus, if we find an assemblage $\sigma_{bc|yz}$ that produces a local behaviour for the set of measurements $\mathcal{E}$, then it also produces a local behaviour for all noisy projective measurements $\Pi_{a|x}(\mu')$, with $\mu' \leq \mu$. This in turn implies that the noisy assemblage $\sigma_{bc|yz}(\mu)$ produces local behaviours for \emph{all} projective measurements $\Pi_{a|x}$. All details of this outline can be found in the supplementary material B.

Putting both ingredients together, if the noisy assemblage $\sigma_{bc|yz}(\mu)$ violates the Tsirelson bound of a steering functional, it is a postquantum assemblage which produces a local (hence quantum) behaviour for all projective measurements. We have used standard iterative optimisation techniques to find such an inequality and assemblage. Again, full details of the approach, including the computational tractability, can be found in the supplementary material D.

Finally, starting from the above example we can use the protocol of \cite{Hirsch13} to construct a new assemblage which is local for all POVMs. The protocol shows that if the qubit assemblage $\sigma_{bc|yz}$ produces local behaviours for all projective measurements, then the qutrit assemblage
\begin{equation}\label{e:POVM assemblage}
\sigma'_{bc|yz} = \tfrac{1}{3}\sigma_{bc|yz}+ \tfrac{2}{3}\tr[\sigma_{bc|yz}] \ket{2}\bra{2}
\end{equation}
produces local behaviours for all (qutrit) POVMs. Moreover, the qutrit assemblage is also postquantum. Indeed, if Alice applies the local filter $F = \ket{0}\bra{0} + \ket{1}\bra{1}$ to the assemblage, she obtains back the original postquantum qubit example. Since a local filter cannot convert a quantum-realizable assemblage into a post-quantum assemblage, the new example is also necessarily postquantum. Full details of the construction can be found in the supplementary material C.

\emph{Example}. We applied the above method in order to find an example of non-trivial post-quantum steering. We first present the qubit assemblage which is local for all projective measurements.  In Fig. 1 we represent graphically the members of the assemblage, denoted $ \sigma_{bc|yz}^*$. This assemblage is real and symmetric under permutation of Bob and Charlie, i.e. $ \sigma_{bc|yz}^* =  \sigma_{cb|zy}^*$. Moreover, it is postquantum: it achieves $\beta = -0.520495$ for an inequality with almost-quantum bound $\beta_\mathcal{\widetilde{Q}} = -0.508417$. Note that the operators $F_{bcyz}$ characterizing the inequality \eqref{ineq}, and more details about $ \sigma_{bc|yz}^*$ can be found in the supplementary material E. Finally, an example which is local for all POVMs follows by applying \eqref{e:POVM assemblage} to the example $\sigma_{bc|yz}^*$.

\begin{figure}[t!]
\includegraphics[width=0.85\columnwidth]{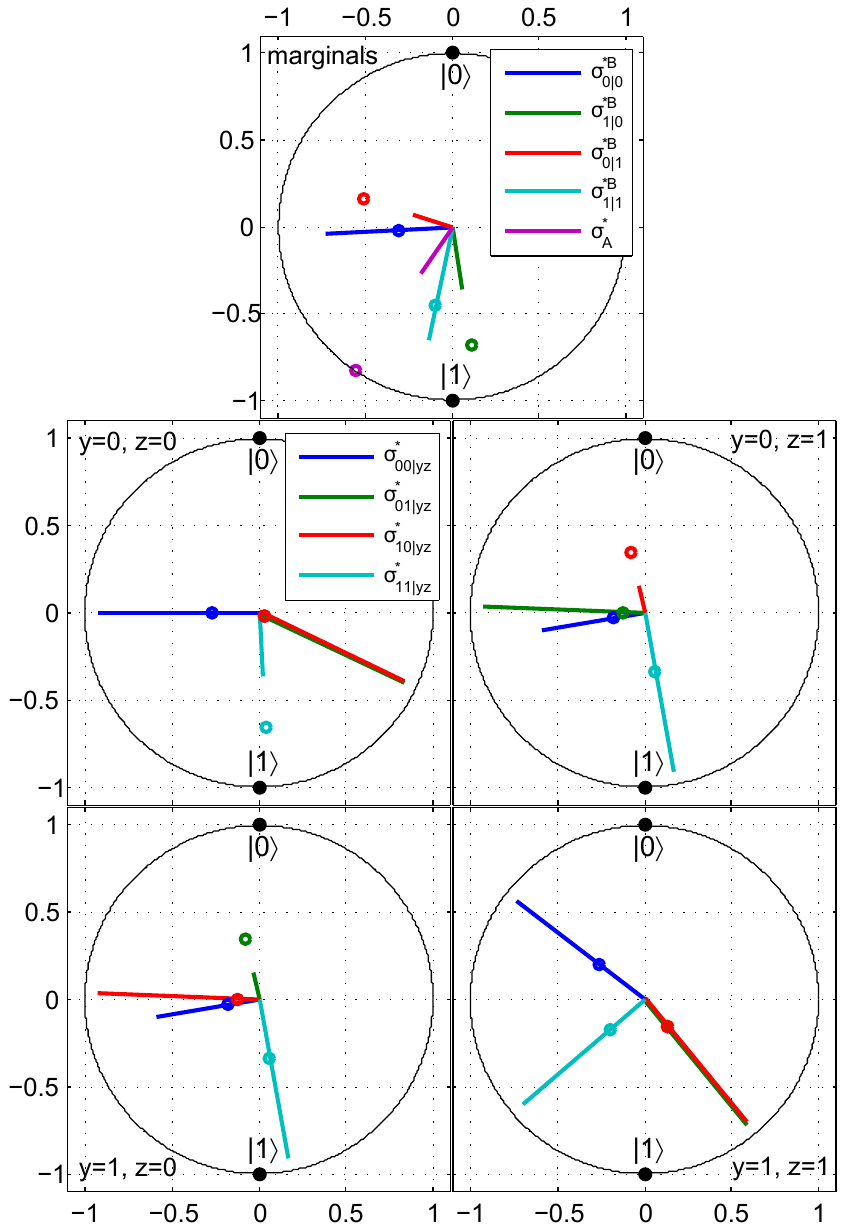}
\caption{Bloch sphere representation of the postquantum assemblage $ \sigma_{bc|yz}^*$. For each pair of settings $y,z$ we represent the four conditional real qubit states $ \sigma_{bc|yz}^*$ in an equator of the Bloch sphere. The normalized state is given by its Bloch vector, while the normalization is indicated by the corresponding circle; more precisely, the distance to the origin corresponds to $p(bc|yz)=\tr(\sigma_{bc|yz}^*)$. The upper figure indicates the marginal states $ \sigma_{b|y}^{B \,*}$, as well as the reduced state $\rho^*_A$. }
\end{figure}

\emph{Discussion.--}Motivated by the development and success of postquantum nonlocality, we have investigated the possibility of extending steering beyond quantum theory. While such an extension is not possible in the bipartite case, we showed explicitly the existence of postquantum steering in the multipartite case. Notably, this represents a genuinely new effect, since postquantum steering does not imply postquantum nonlocality. Hence the use of postquantum resources can only be witnessed by looking at the assemblage, but is not apparent at the level of the nonlocal behaviours it can produce.

An interesting aspect of this work is to highlight a fundamental difference between bipartite and multipartite quantum correlations. This goes alongside previous findings \cite{acinO,toni}. For instance, in the case of nonlocality it was shown that a natural extension of Gleason's theorem is possible in the bipartite case, but fails for multipartite systems \cite{acinO}. In the context of entanglement theory, every pure bipartite entangled state admits a canonical form (Schmidt decomposition), however the situation turns out to be more complex in the multipartite case \cite{toni}. It would be very interesting to understand whether the above observations are intimately related to each other and to the existence of postquantum steering.

Our work raises several questions. For instance, it would also be interesting to find further examples of postquantum steering, and understand how generic the phenomenon is. Moreover, given the strong information-theoretic power of certain postquantum nonlocal correlations, it would be relevant to investigate what can be achieved using postquantum steering. In particular, can post-quantum steering enhance protocols involving quantum information, for instance better quantum teleportation or remote state preparation?

\emph{Acknowledgements.}---We thank Flavien Hirsch for discussions, and acknowledge financial support from: the Swiss National Science Foundation (grant PP00P2\_138917 and Starting Grant DIAQ); SEFRI (COST action MP1006); the EU SIQS; the Beatriu de Pin\'os fellowship (BP-DGR 2013); EPSRC grant DIQIP; ERC AdG NLST; ERC CoG QITBOX; the J\'anos Bolyai Programme; and the OTKA grant (K111734).

\section{Appendix A: Lower bounding the steering Tsirelson bound through a relaxation of the set of quantum assemblages}\label{ap:sdp}

In this appendix we present the details of a relaxation of the set of quantum assemblages, which we call `almost quantum' and denote by $\mathcal{\widetilde{Q}}$ \cite{TorNavVer14}. The name comes from its close relation to the definition of set of almost quantum correlations, which is also characterised by an SDP \cite{NPA}. It is this relaxation which will allow us to compute lower bounds on the Tsirelson bound of any steering functional.

Similarly to the NPA hierarchy of Ref. \cite{NPA}, consider a moment matrix $\Gamma$ whose rows and columns are labelled by the `words' from the following set:

\begin{widetext}
\begin{equation}\label{theset}
\mathcal{S} := \{ \emptyset \} \cup \left\{ (b|y) \right\}_{\substack{b=1:k_B-1 \\ y=1:m_b}} \cup \left\{ (c|z) \right\}_{\substack{c=1:k_C-1 \\ z=1:m_c}} \cup \left\{ (bc|yz) \right\}_{\substack{b=1:k_B-1, \, c=1:k_C-1\\ y=1:m_b, \, z=1:m_c}},
\end{equation}
\end{widetext}
where $m_b$ denotes the possible measurement choices by Bob, each with $k_B$ number of outcomes (and similarly for Charlie). In the NPA hierarchy, a matrix with such labels is the object of study for the 1+AB level, where some elements of the matrix are related to the values of a conditional probability distribution $p(bc|yz)$ and its marginals. In our case, however, each of the elements of $\Gamma$ corresponds to a conditional state prepared on Alice's side, in a way that we make explicit below.

The elements of the first row of $\Gamma$ are set as follows: 

\begin{align}
\label{frr_i} \Gamma(\emptyset, \emptyset) &:= \rho_\mathrm{A}, \\
\Gamma(\emptyset, b|y) &:= \sigma_{b|y}, \\
\Gamma(\emptyset, c|z) &:= \sigma_{c|z}, \\
\label{frr_f}\Gamma(\emptyset, bc|yz) &:= \sigma_{bc|yz},
\end{align}
where the reduced states are as in Eqs (4b) to (4c) of the main text.

Once such an identification is done, further constraints are imposed between the elements of $\Gamma$ to enforce some quantum-like properties on the assemblage. In order to make it clearer to the reader, we present first the relation between $\Gamma$ and quantum assemblages, from which the extra constraints on the moment matrix will hopefully arise naturally.

A quantum assemblage arises by Bob and Charlie performing measurements on their share of a tripartite quantum system $\rho_{\mathrm{ABC}}$. Let $E_{b|y}$ and $E_{c|z}$ be the POVM elements. Note that we can assume them to be projectors, since in principle we do not impose any constraints on the dimensions of Bob and Charlie's subsystems. The assemblage then arises as:
\begin{align}
\label{fr_i}\rho_\mathrm{A} &= \tr_{\rB\rC}\left( \rho_\mathrm{ABC} \right), \\
\sigma_{b|y} &= \tr_{\rB\rC} \left( \id_\mathrm{A} \, E_{b|y} \, \id_\mathrm{C} \, \rho_\mathrm{ABC} \right),\\
\sigma_{c|z} &= \tr_{\rB\rC} \left( \id_\mathrm{A} \, \id_\mathrm{B} \, E_{c|z} \, \rho_\mathrm{ABC} \right),\\
\label{fr_f}\sigma_{bc|yz} &= \tr_{\rB\rC} \left( \id_\mathrm{A} \, E_{b|y} \, E_{c|z} \, \rho_\mathrm{ABC} \right).
\end{align}
Note that we are using the commutativity paradigm, where we do not require that the measurements be of the form $\id_\mathrm{A} \otimes E_{b|y} \otimes E_{c|z}$, but rather demand that $\left[ E_{b|y}, E_{c|z} \right]= 0$ for all $b,c,y,z$. The tensor product type of measurements is just a particular case of the general form of the latter.

Now consider the moment matrix again. To each of its entries we associate the following element:
\begin{align}
\Gamma(v,w) &= \tr\left( \op_v^\dagger \, \op_w \, \rho_\mathrm{ABC} \right), \\
\mathrm{where} \quad \op_\emptyset &= \id \\
\op_{b|y} &= \id_\mathrm{A} \, E_{b|y} \, \id_\mathrm{C}, \\
\op_{c|z} &= \id_\mathrm{A} \, \id_\mathrm{B} \, E_{c|z}, \\
\op_{bc|yz} &= \id_\mathrm{A} \, E_{b|y} \, E_{c|z}.
\end{align}
It is clear to see that the elements of the first row $\Gamma(\emptyset,v)$ satisfy eq. (\ref{fr_i}) to (\ref{fr_f}) for all $v$. In addition, the commutation relations between the measurement operators of Bob and Charlie also impose that:
\begin{align}
\label{p_i}\Gamma(v,v) &= \Gamma(\emptyset ,v),\\
\Gamma(v,w) &= \Gamma(w, v), \quad \mathrm{whenever} \, [\op_v,\op_w]=0,
\end{align}
and constraints of the type:
\begin{align}
\Gamma(b|y,bc|yz) &= \Gamma(\emptyset, bc|yz),\\
\Gamma(b|y,bc|yz) &= \Gamma(b|y,c|z),\\
\Gamma(bc|yz,bc^\prime|yz^\prime) &= \Gamma(bc|yz, c^\prime|z^\prime).
\end{align}
Note that these constraints are the ones imposed on the matrix moment of the 1+AB level of the NPA hierarchy. In our case, however, the elements of $\Gamma$ are matrices instead of numbers, and hence some specific properties also arise. These are of the type:
\begin{align}
\Gamma(b|y, b^\prime|y^\prime) &= \Gamma(b^\prime|y^\prime, b|y)^\dagger,\\
\Gamma(bc|yz, b^\prime c|y^\prime z) &= \Gamma(b^\prime c|y^\prime z, bc|yz)^\dagger,\\
\label{p_f}\Gamma(bc|yz, b^\prime|y^\prime) &= \Gamma(b^\prime c|y^\prime z, b|y)^\dagger.
\end{align}
Finally, note that such a $\Gamma$ is hermitian and positive semidefinite.

The idea now is, given a general assemblage $\left\{ \sigma_{bc|yz} \right\}_{bcyz}$ check whether there exists a PSD moment matrix $\Gamma$ whose first row relates to the assemblage via eq. (\ref{fr_i}) to (\ref{fr_f}), and that satisfies properties (\ref{p_i}) to (\ref{p_f}). This is a well defined semidefinite program, and when it is feasible the assemblage belongs to $\mathcal{\widetilde{Q}}$. Since every quantum assemblage satisfies the properties, such an SDP is always feasible for quantum inputs, hence every quantum assemblage belongs to $\mathcal{\widetilde{Q}}$. Note that the converse may not always be true.

We use this set $\mathcal{\widetilde{Q}}$ to find bounds on the Tsirelson bound of a steering functional. Since $\mathcal{\widetilde{Q}}$ may contain post- quantum assemblages, a lower bound on $\beta_\mathcal{Q}$ is obtained by finding the minimum value of the functional over $\mathcal{\widetilde{Q}}$, which is itself an SDP:
\begin{align}
\mathrm{minimise} \quad & \tr \left( \sum_{bcyz} F_{bcyz} \, \sigma_{bc|yz} \right) \\
\mathrm{such ~that} \quad & \left\{ \sigma_{bc|yz} \right\}_{bcyz} \in \mathcal{\widetilde{Q}}.
\end{align}

For the scope of this work we only need a bound on $\beta_\mathcal{Q}$, whose violation ensures that the assemblage is post-quantum. We do not need to study different optimal bounds on $\beta_\mathcal{Q}$ or other relaxations of the quantum set of assemblages. For the reader interested in SDPs, however, that may also be a valid question and we comment on it in what follows.

A natural step towards studying different relaxations of the quantum set goes in spirit with the NPA hierarchy, similar to the idea by Pusey for bipartite steering scenarios \cite{pusey13}. One could consider then a hierarchy of moment matrices $\Gamma_n$, where $n$ relates to length of the words in the set (\ref{theset}), which is now allowed to contain elements of the form $(b_1 \ldots b_jc_1 \ldots c_k | y_1 \ldots y_jz_1 \ldots z_k)$. In the case of quantum assemblages, such indices would relate to the following:
\begin{widetext}
\begin{align}
&\Gamma(b_1 \ldots b_{j_1}c_1 \ldots c_{k_1} | y_1 \ldots y_{j_1} z_1 \ldots z_{k_1}, b'_1 \ldots b'_{j_2}c'_1 \ldots c'_{k_2} | y'_1 \ldots y'_{j_2} z'_1 \ldots z'_{k_2}) = \\
&  \tr \left(\id_\mathrm{A} \, E^\dagger_{b_{j_1} | y_{j_1}} \ldots E^\dagger_{b_{1} | y_{1}} \, E^\dagger_{c_{k_1} |z_{k_1}} \ldots E^\dagger_{c_{1} | z_{1}} \, E_{b'_{1} | y'_{1}} \ldots E_{b'_{j_2} | y'_{j_2}}  \, E_{c'_{1} | z'_{1}} \ldots E_{c'_{k_2} | z'_{k_2}} \, \rho_{\mathrm{ABC}} \right).\nonumber
\end{align}
\end{widetext}

From the commutations relations between Bob and Charlie's measurements arise different constrains that $\Gamma_n$ is asked to satisfy. Note that the longer the words in $\mathcal{S}_n$ are, the more the properties that the moment matrix should satisfy. For each $n$, testing whether those properties are satisfied when some elements of the first row are set to be the conditional states on Alice's side (eq. (\ref{frr_i}) to (\ref{frr_f})) is an SDP, and feasibility of level $n$ implies feasibility of level $m<n$. This last statement follows from the fact that every word in $\mathcal{S}_m$ is a word on $\mathcal{S}_n$, hence the constraints imposed in level $m<n$ are just a subset of those imposed in level $n$. Note also that when the input is a quantum assemblage, the SDP is feasible for any level $n$ by definition.

Denote by $\mathcal{Q}_n$ the set of assemblages which satisfy the conditions of the level $n$ SDP. Then, the following SDPs define a sequence of lower bounds to the Tsirelson bound of a steering functional:
\begin{align}
\mathrm{minimise} \quad & \beta_{\mathcal{Q}_n} = \tr \left( \sum_{bcyz} F_{bcyz} \, \sigma_{bc|yz} \right) \\
\mathrm{such~that} \quad & \left\{ \sigma_{bc|yz} \right\}_{bcyz} \in \mathcal{Q}_n.
\end{align}

By definition, these lower bounds satisfy $\beta_{\mathcal{Q}_m} \leq  \beta_{\mathcal{Q}_n}$ whenever $m<n$.

\section{Appendix B: Details for constructing a local model for all projective measurements}
The second ingredient in our construction is a method for constructing qubit assemblages $\sigma_{bc|yz}$ that always give rise to quantum-realizable behaviours, that is $p(abc|xyz) = \tr_\mathrm{A}(\Pi_{a|x}\sigma_{bc|yz})$ admits a quantum realization for any possible projective measurement $\Pi_{a|x}$ performed by Alice.

To simplify the problem, we restrict to assemblages of real-valued qubit states, i.e. such that all conditional states lie in the $x$-$z$ plane of the Bloch sphere. It follows that we need only consider projective measurements in the $x$-$z$ plane (since the $y$ component identically vanishes), parametrised by
\begin{equation}\label{meas_par}
\Pi_{a|\theta} = \frac{\left( \openone + (-1)^a (\cos(\theta) \, X + \sin(\theta) \, Z) \right) }{2},
\end{equation}
where $\theta \in [0, \pi)$, $a = 0,1$, and $X$ and $Z$ denote the corresponding Pauli matrices (the range $\theta \in [\pi, 2\pi)$ comes from Alice relabelling her outcome $0 \leftrightarrow 1$).

Ensuring that the behaviour $p(abc| \theta yz) = \tr_\mathrm{A}(\Pi_{a|\theta}\sigma_{bc|yz})$ admits a quantum realization is a difficult problem in general. Instead, we demand that $p(abc| \theta yz) $ is Bell local, that is, admits a decomposition of the form
\begin{equation}
p(abc| \theta yz) = \int dÊ\lambda \pi(\lambda) p(a|\theta \lambda) p(b|y \lambda) p(c|z \lambda)
\end{equation}
where $\lambda$ denotes a shared local variable, distributed according to the density $\pi(\lambda)$ \cite{review}. Indeed, any behaviour which is Bell local is also realizable in quantum theory.

Next, we construct an assemblage admitting a local model by adapting the ideas of Ref. \cite{Bowles2015}. Specifically, consider the set of four measurements $\mathcal{E} = \{\Pi_{a|\theta_x} \}_{ax}$, where $\theta_x = x\pi/4$ and $x = 0,\ldots, 3$. Next take an assemblage $\sigma_{bc|yz}$ such that the behaviour $p(abc| \theta_x yz)= \tr_\mathrm{A}(\Pi_{a|\theta_x} \sigma_{bc|yz}) $ is local. For a given assemblage, this can be easily verified using linear programing \cite{review}.

Following Ref. \cite{Bowles2015}, this implies that $\sigma_{bc|yz}$ is local for all noisy two-outcome projective measurement in the $x$-$z$ plane:
\begin{align}
\Pi_{a|\theta}(\mu) &= \mu \, \Pi_{a|\theta} + (1-\mu) \, \openone/2
\end{align}
whenever $\mu \leq \cos(\pi/8)$. This follows from the fact that any operator $\Pi_{a|\theta}(\mu)$ can be written as a convex combination of the 4 measurements in $\mathcal{E}$. That is, for all $a$ and $\theta$, we can find coefficients $c_{a'x} \geq 0$ with $\sum_{a'x} c_{a'x} = 1$ such that
\begin{equation} \Pi_{a|\theta}(\mu) = \sum_{a'=0}^1 \sum_{x=0}^3 c_{a'x} \Pi_{a'|\theta_x}.
\end{equation}
This can be seen geometrically; the Bloch vectors of the 8 POVM elements in $\mathcal{E}$ form an octagon in the $x$-$z$ plane of the Bloch sphere, and $\cos(\pi/8)$ is the radius of the largest circle that fits inside this octagon.

Next, we note the following equality
\begin{equation}
 \tr\left( \Pi_{a|\theta}(\mu) \, \sigma_{bc|yz} \right) = \tr\left( \Pi_{a|\theta} \, \sigma_{bc|yz}(\mu) \right)
\end{equation}
where
\begin{equation}\label{noisy ass}
\sigma_{bc|yz} (\mu)= \mu \, \sigma_{bc|yz} + (1-\mu) \, \tr\left( \sigma_{bc|yz} \right) \openone/2.
\end{equation}
That is, the statistics of noisy measurements on the assemblage $\sigma_{bc|yz}$ perfectly match the statistics of projective measurements on the noisy assemblage $\sigma_{bc|yz}(\mu)$. Hence, if $\sigma_{bc|yz}$ is local for the four measurements in $\mathcal{E}$, then $\sigma_{bc|yz}(\mu)$ is local for all projective measurements when $\mu \leq \cos(\pi/8)$. We have thus constructed an assemblage, $\sigma_{bc|yz}(\mu)$, which gives rise to behaviours which are Bell local, hence admit a quantum realization.

\section{Appendix C: Details of constructing a POVM qutrit model from a projective qubit model}
In this appendix we outline how given an qubit assemblage which is local for all projective measurements, we can find a qutrit assemblage which is local for all POVMs. The idea for this construction comes from Ref. \cite{Hirsch13}, which we apply to our scenario.

First, assume that an example $\sigma_{bc|yz}$ is given which is a collection of qubits which produces a local behaviour for all projective measurements. Let us consider that in fact $\sigma_{bc|yz}$ is a collection of qutrits, which have support only on the qubit ($\ket{0}$, $\ket{1}$) subspace. It follows then that the assemblage also produces local behaviours for all dichotomic projective qutrit measurements. This follows immediately, since any dichotomic projective qutrit measurement, when restricted to the qubit subspace (the only subspace where Alice's states have support) is a noisy dichotomic qubit measurement. This is a convex combination of projective measurements, and hence can be covered by the local model; see \cite{Hirsch13} for more details.

Now, we apply protocol 2 of \cite{Hirsch13}: Without loss of generality we can restrict to POVMs with each element $E_a = \alpha_a \Pi_a$, for $\Pi_a$ a projector, and $\sum_a \alpha_a = 3$. Alice chooses the projector $\Pi_\alpha$ with probability $\alpha_a/3$. She simulates the dichotomic measurement $\{\Pi_a,\openone-\Pi_a\}$ on $\sigma_{bc|yz}$. If the outcome corresponds to $\Pi_a$, she gives as outcome $a$. Otherwise, she gives as outcome $a'$ with probability $\bra{2} E_{a'} \ket{2}$. The totally probability for Alice to give as outcome $a$ is 
\begin{align*}
\frac{\alpha_a}{3}\tr[\Pi_a \sigma_{bc|yz}] + \sum_{a'}\frac{\alpha_{a'}}{3}\tr[(\openone-\Pi_{a'})\sigma_{bc|yz}\bra{2}E_a \ket{2} \nonumber \\
= \frac{1}{3} \tr[E_a\sigma_{bc|yz}] + \frac{2}{3}\tr[\sigma_{bc|yz}]\bra{2}E_a\ket{2}
\end{align*}
which is the same as would be obtained by measuring the POVM $E_a$ on the assemblage $\sigma'_{bc|yz}$, where
\begin{equation}
\sigma'_{bc|yz} = \frac{1}{3}\sigma_{bc|yz} + \frac{2}{3}\tr[\sigma_{bc|yz}]\ket{2}\bra{2}.
\end{equation}
Thus, if Alice has a local model for all dichotomic projective measurements on $\sigma_{bc|yz}$ then she has a local model for all POVMs on $\sigma'_{bc|yz}$.

Finally we note that if Alice instead performs a local filter $F=\ket{0}\bra{0}+\ket{1}\bra{1}$, then the assemblage after filtering $F\sigma'_{bc|yz}F^\dagger/\tr[F\sigma_{bc|yz}F^\dagger] = \sigma_{bc|yz}$ is the original qubit assemblage. Local filtering cannot convert an assemblage with a quantum realization to one without a quantum realization (otherwise quantum theory would not be closed under filtering). As a corollary, if the assemblage after filtering has no quantum realization, then the assemblage beforehand also cannot have a quantum realization. Thus, every example of a post-quantum qubit assemblage which has a local model for all projective measurements immediately leads to a post-quantum qutrit assemblage which has a local model for all POVMs using the above construction.

\section{Appendix D: Details of numerical search algorithm}
In this appendix we give details about how all the ingredients can be put together to search for non-trivial examples of post-quantum steering. We recall that we can restrict to searching for examples which are local for all projective measurements, since the final step of constructing an example which is local for all POVMs then follows analytically.

Our task is to find a steering functional with elements $F_{bcyz}$ and almost quantum bound $\beta_\mathcal{\widetilde{Q}}$, and a real-valued assemblage $\sigma_{bc|yz}$ such that the following two constraints are satisfied
\begin{align}\label{e:final constraints}
& \tr\sum_{bcyz} F_{bcyz} \, \sigma_{bc|yz}^*  < \beta_\mathcal{\widetilde{Q}}  \\ \label{cond2}
& p(abc|\theta_xyz) = \tr\big( \Pi_{a|\theta_x} \, \sigma_{bc|yz} \big) \text{ is local} \quad \forall \Pi_{a|\theta_x} \in \mathcal{E} \nonumber \\
\end{align}
where $ \sigma_{bc|yz}^*  =  \sigma_{bc|yz}(\mu = \cos(\pi/8))$ as defined in Eq. \eqref{noisy ass}. The noisy assemblage $\sigma_{bc|yz}^*$ is then the example we are looking for.
From condition \eqref{e:final constraints} it follows that $\sigma_{bc|yz}^*$ does not admit a quantum realization. From condition \eqref{cond2} it follows that the statistics of any projective measurements on $\sigma_{bc|yz}^*$ gives rise to a Bell local distribution, hence realizable in quantum theory.

In practice, certifying whether there exists an assemblage $\sigma_{bc|yz}$ which satisfies the conditions \eqref{e:final constraints} and \eqref{cond2} for a given set of operators $F_{bcyz}$ can be checked by solving a semidefinite program (SDP). More precisely, we treat the left hand side of condition \eqref{e:final constraints} as the objective function, while keeping \eqref{cond2} as a constraint, i.e. we solve
\begin{align}
\beta = \min_{\sigma^*_{bc|yz}} &\quad\tr\sum_{bcyz} F_{bcyz}\, \sigma_{bc|yz}^* \nonumber\\
\text{s.t.} &\quad \tr\big( \Pi_{a|\theta_x} \, \sigma_{bc|yz} \big) \text{ is local} \quad \forall \Pi_{a|\theta_x} \in \mathcal{E}
\end{align}
The last part of the problem is to judiciously choose the operators $F_{bcyz}$. Here we employed the following method: (i) we generate randomly real-valued operators $F_{bcyz}$ and calculate the lower bound on the Tsirelson bound $\beta_\mathcal{\widetilde{Q}}$, itself an SDP. (ii) we solve the SDP described above with $\mu = 1$. If $\beta \geq \beta_\mathcal{\widetilde{Q}}$ we abort, and restart. If $\beta < \beta_\mathcal{\widetilde{Q}}$ we calculate $\mu$ such that $\tr\sum_{bcyz} F_{bcyz}\sigma_{bc|yz}(\mu) =  \beta_\mathcal{\widetilde{Q}}$. If $\mu \leq \cos(\pi/8)$ we have the desired example. Otherwise we return to the beginning, applying standard gradient descent methods in order to generate a new set of operators $F_{bcyz}$. This method was implemented in {\sc matlab} using {\sc cvx}  \cite{cvx}.

\begin{table}[t!]
\begin{tabular}{ll}
$\rho_\mathrm{A}^* = \left( \begin{array}{rr}0.3666 &-0.0896 \\-0.0896 &0.6334
\end{array} \right)$
 &
 $\sigma_{00|00}^* = \left( \begin{array}{rr}0.1360 &-0.1257 \\ -0.1257 &0.1360
\end{array} \right)$
 \\
$\sigma_{0|0}^\mathrm{B \, *} = \left( \begin{array}{rr}0.1464 &-0.1114  \\-0.1114 &0.1600
\end{array} \right)$
&
$\sigma_{00|10}^* = \left( \begin{array}{rr}0.0803 &-0.0523  \\ -0.0523 &0.0982
\end{array} \right)$
\\
$\sigma_{0|1}^\mathrm{B \, *} = \left( \begin{array}{rr}0.2851  &-0.0586  \\-0.0586  &0.2473
\end{array} \right)$
&
$\sigma_{00|11}^* = \left( \begin{array}{rr}0.2555  &-0.1192  \\-0.1192  &0.0709
\end{array} \right)$
  \end{tabular}
  \caption{Example of post-quantum assemblage that cannot lead to post-quantum nonlocality (for arbitrary projective measurements).   \label{tab:ass}}

\begin{tabular}{ll}
$ F_A = \left( \begin{array}{rr}1.4622 &0.1773 \\
0.1773 &-0.4622
\end{array} \right)$
 &
$F_{00} = \left( \begin{array}{cc}-0.1948 &0.5653  \\
0.5653 &-0.7229
\end{array} \right)$
 \\
$F^\mathrm{B}_{0} = \left( \begin{array}{cc}-0.2894 &0.2468 \\
0.2468 &0.9767
\end{array} \right)$
&
$F_{10} = \left( \begin{array}{cc} 0.5482 &-0.4270  \\
-0.4270 &-0.8690
\end{array} \right)$
\\
$F^\mathrm{B}_{1}  = \left( \begin{array}{cc}-1.0943 &-0.4673 \\
-0.4673 &0.0648
\end{array} \right)$
&
$F_{11} = \left( \begin{array}{cc}0.2875 & 1.0320  \\
1.0320 &0.9182
\end{array} \right)$
  \end{tabular}
 \caption{Operators defining an inequality of the form of Eq. (5) of the main text, which witnesses the fact that the assemblage  given in Table \ref{tab:ass} is post-quantum.} \label{tab:ineq}
 \end{table}

\section{Appendix E: Details about example of post-quantum steering}\label{ap:2}

We give here the details concerning the example of a post-quantum qubit steering without post-quantum nonlocality for projective measurements. This example then leads to the qutrit example without post-quantum nonlocality for all POVMs. 

The qubit assemblage $\sigma_{bc|yz}^*$ is given explicitly in Table I; in the main text, we represented graphically the assemblage in Fig. 1. Note that we present $ \sigma_{bc|yz}^*$ in a minimal representation, using the no-signalling and normalization conditions (4) (see main text), and symmetry under permutation of Bob and Charlie, i.e. $ \sigma_{bc|yz}^* =  \sigma_{cb|zy}^*$.

Moreover, in Table II, we give the operators $F_{bcyz}$ for constructing the steering functional of Eq. (5) of the main text. These operators are also given in minimal representation, where
$F_A = \sum_{yz} F_{11yz}$, $F^\mathrm{B}_{y} = \sum_{bz} (-1)^b F_{b1yz}$, $F^\mathrm{C}_{z} = \sum_{cy} (-1)^c F_{1cyz}$, and $F_{yz} = \sum_{bc} (-1)^{b+c} F_{bcyz}$. The quantity in Eq. (5) is then calculated as follows:
\begin{equation}
\beta = \tr \left(  F_A \rho_A + \sum_y F^\mathrm{B}_{y} \sigma_{0|y}^\mathrm{B }
+ \sum_z F^\mathrm{C}_{z} \sigma_{0|z}^\mathrm{C }  + \sum_{yz} F_{yz} \sigma_{00|yz}  \right).
\end{equation}

To obtain the final example of a non-trivial post-quantum assemblage we apply the procedure of Appendix C to convert the above qubit assemblage into a qutrit assemblage, which is guaranteed to produce local behaviours for all POVMs whilst still being post-quantum.


\begin{thebibliography}{99}
\bibitem{footnote1} Whenever it will cause no confusion, for ease of presentation we will use $\sigma_{bc|yz}$ to refer to the entire assemblage, i.e. the set $\{\sigma_{bc|yz}\}_{bcyz}$.
\bibitem{footnote3} Geometrically, a shrunken Bloch ball can always be contained inside a polyhedron which is itself contained inside the Bloch ball.
\end{thebibliography}
\end{document}